\documentclass{JHEP3}

\usepackage{amsfonts}
\usepackage{amsmath}
\usepackage{amssymb}
\usepackage{amsmath,amssymb,epsfig}
\numberwithin{equation}{section}
\usepackage{cite}

\newcommand{\be}{\begin{equation}}
\newcommand{\bea}{\begin{eqnarray}}
\newcommand{\eea}{\end{eqnarray}}
\newcommand{\ba}{\begin{array}}
\newcommand{\ea}{\end{array}}
\newcommand{\ee}{\end{equation}}

\title{$R$-symmetric Gauge Mediation With Fayet-Iliopoulos Term}

\author{Mingxing Luo and Sibo Zheng \\
Zhejiang Institute of Modern Physics, Department of Physics,\\
Zhejiang University, Hangzhou 310027, P. R. China.\\
 E-mail: \email{luo@zimp.zju.edu.cn}, \email{sibozheng.zju@gmail.com}}

\abstract{We have studied $R$-symmetrc gauge mediation models with
Fayet-Iliopoulos terms. We give a concrete example of hidden
sector with an $U(1)_H$ gauge theory and a Fayet-Iliopoulos term,
which can induce distinctive soft terms in the visible sector, and
help solving fine tuning problems in models of $R$-symmetric gauge
mediation.}

\begin{document}
\section{Introduction}
Of all the candidates to stabilize the hierarchy between the weak
and Plank scales, supersymmetry seems to be the most plausible and
predictive at LHC. Supersymmetry is broken in some hidden sector
and mediated to the visible sector via gravity or gauge
interactions \cite{Giudice,Luty}.
Such mechanisms induce the necessary soft terms in supersymmetric
Standard Model (SSM) for which to be phenomenoloically viable. In
models with direct gauge mediation, a number of hidden models have
been successfully constructed \cite{GAUGE MEDIATION}. There exist
now a general framework to calculate the soft masses \cite{MSS}.

Recently, a new class of gauge mediation models has been proposed
\cite{Poppitz,Poppitz07}, in which $R$ symmetry is retained in
this class of models. In usual gauge mediation models, $R$
symmetry has to be spontaneously broken in order to generate
Marojana gaugino massses.  By adding suitable $C$-parity chiral
fields in adjoint representations, the gauginos can acquire Dirac
masses. Phenomenologies are rather distinctive in these kind of
models \cite{Poppitz07}. Unfortunately, some of the usual flavor
problems persist in these models and fine tunings are needed.

In this paper, we will first have a close look at hidden sectors
with $R$-symmetry and SUSY-breaking. It has been long understood
that these two issues are closely connected to each other
\cite{Nelson93}. The $R$-symmetry can be broken at the tree level.
If it is not broken at the tree level, it can still be
spontaneously broken due to quantum corrections in general O
$^{'}$Raifeartaigh (OR) model, provided that the $R$-charges of
superfields in the superpotential take values different from 0 and
2 \cite{Shih}. This implies that the general form of
superpotential is determined as $W=X^{n}f_{n}(\Phi)$, where
$X,\Phi$ are $R$-charges 2 and 0 respectively, $f_{n}(\Phi)$ are
polynomial of $\Phi$ constrained only by the renormalization of
the theory. In section II, we will deform these models by
including a Fayet-Iliopoulos (FI) term. And one sees that the
$R$-symmetry can still be preserved.

In section III, we will construct a hidden sector with a FI term
to realize the scenario of $R$-symmetric gauge mediation. Embedded
into SUSY theories via an $U(1)$ group, the phenomenological
implications of the FI term have been extensively discussed in
gauge and anomaly mediation\footnote {Applications in other
fields, such as extra dimensions, cosmology and string theory, are
beyond present discussions.}. In gauge mediation scheme, an $U(1)$
FI term can be used to spontaneously break supersymmetry, while
$R$-symmetry is usually unbroken as proposed in \cite{Nelson93}.
In $N=2$ models, this can naturally generate Dirac gaugino
masses\cite{0206096}. In \cite{0705}, an FI term is employed for
$N=1$ models with Majorana gaugino masses, in which $R$-symmetry
is spontaneously broken by gaugino condensation in a
strong-coupled Yang-Mills hidden sector. FI terms can also be
applied in the construction of SUSY grand unification theories
(GUT). If the role of messenger sector are replaced by an $U(1)$
with FI term, there will be less modifications to RG running of SM
gauge couplings. A GUT can be relatively easier realized
\cite{9806430}. Moreover, doublet-triplet splitting problem in
high rank $SU(n)$ SUSY theories that include $SU(5)$ GUT can be
solved by introducing a new mass scale carried by FI term
\cite{9611277}. In anomaly mediation scheme, introducing a FI term
can help to solve the tachyonic slepton problem, or even
accommodate neutrino masses when suitable $U(1)$ charges are
chosen \cite{0302209,0202101}.

In section IV, we calculate the soft masses, and $\mu/B\mu$ terms.
The model preserves most distinctive features of $R$-symmetric
gauge mediation. However, we can allievate the fine tunings and
the $\mu/B\mu$ problem. To summarize, the model has the following
features:
\begin{itemize}
    \item $R$-symmetry is conserved in SUSY-breaking and visible sectors.
    \item There are no $A$ terms,
and adjoint fields $\Phi_{r}$ (r=1,2,3) that combined with
gauginos have to be added to construct Dirac gauginos.
    \item There is an unbroken $U(1)_{H}$ gauge theory in the hidden sector.
The SUSY-breaking effects contain the contributions of $F$- and
$D$-terms at meantime.
    \item The negative sfermions masses squared coming from the $D$-term may solve the fine tuning problems in
$R$-symmetric gauge mediation with only $F$-term induced visible
effects \cite{Poppitz}. The Dirac gaugino masses can be heavier
than the sfermions masses.
    \item The little hierarchy between $\mu$ and $B\mu$ terms can also be obtained by adjusting the $D$-term.
\end{itemize}
We expect that some of phenomenologies in the visible sector are
model dependent. It would be interesting to develop a general
framework to distinguish the general characters in $R$-symmetric
gauge mediation, as done in \cite{MSS}. On the other hand, the
$\mu/B\mu$ problem should also be discussed in depth to find a
more economical mechanism. Finally, we conclude in section V with
discussions.

\section{SUSY-breaking Sectors with $R$-symmetry}

According to \cite{Shih}, the $R$-symmetry is maintained only if
that the R charges of superfields in the superpotential are either
0 or 2. Otherwise, radiative corrections will spontaneously break
the $R$-symmetry even if it is conserved at tree level. This
implies that the general form of superpotential with $R$-symmetry
reads,
\begin{eqnarray}
W=X^{n}f_{n}(\Phi_{m}),
\end{eqnarray}
where $R(X^{n})=2$ and $ R(\Phi_{m})=0$, $n,m$ denote different
superfields of $X$ and $\Phi$, respectively, repeated $n$ implies
summation. $f_{n}(\Phi)$ are $n$ polynomial of $\Phi$ constrained
only by the renormalization of the theory. The supersymmetry
preserving vacua is given by vanishing F terms:
\begin{eqnarray}
F_{n}=f_{n}(\Phi_{m})=0,~~~~~F_{m}=X^{n}\partial_{m}f_{n}(\Phi_{m})=0
\end{eqnarray}

$F_{m}=0$ can easily be satisfied by simply taking all $X^{n}=0$.
In cases with $n>m$, generally $F_{n}=0$ can not be simultaneously
satisfied, thus leading to supersymmetry spontaneously broken. The
$R$-symmetry is preserved at the origin of the moduli space. In
cases with $n\leq m$, there are two possibilities. In a generally
renormalizable theory, $f_{n}(\Phi_{m})$ always assumes the
following form,
\begin{eqnarray}
f_{n}(\Phi_{m})=k_{n}+M_n^m \Phi_{m}+ \lambda_n^{m m^{'}}
\Phi_{m}\Phi_{m^{'}}
\end{eqnarray}
If all $k_{n}=0$, we expect the supersymmetry and $R$ symmetry are
both unbroken at the origin of the moduli space. On the other
hand, if some $k_{n}\neq 0$ as in the ordinary OR model, it is
then possible to obtain SUSY-breaking models with $R$ symmetry.
For example, the ISS model constructed in \cite{ISS} belongs to
this type.

If there is an $U(1)_H$ gauge interaction in the hidden sector,
one can also have a FI term in principle. This yields the
mechanism of $D$-term scenario, in addition to the $F$-term
scenario, to induce SUSY-breaking. Generically, a hidden sector
with a $U(1)_{H}$ gauge can have the following potential,
\begin{eqnarray}
V=F_{m}^{2}+F_{n}^{2}+\frac{D^{2}}{2g^{2}},~~~~
D=g^{2}\left(\xi+g^{2}q_{i}(\mid\Phi_{i}\mid^{2}-\mid\Phi_{i}'\mid^{2})\right)
\end{eqnarray}
where $g$ is the $U(1)_{H}$ gauge coupling, $q_{i}$ are the
$U(1)_{H}$ charges of $\Phi_{i}$. Without $k_{n}$ terms, one
obviously cannot have $F=0$ and $D=0$ in the same time. The
supersymmetry is thus broken spontaneously. In the specific case
that the absolute minimum of the potential is at $\Phi=0$, one
would have $V=g^{2}\xi^{2}/2$ and the SUSY-breaking comes only
from the $D$-term.

In the case that there are non-zero $k_{n}$'s, the SUSY-breaking
will come both from $F$- and $D$-term, if we can arrange the
parameters in the model such that the absolute minimum of the
potential is still at $\Phi=0$. In this case, the $R$ symmetry is
unbroken with a minima $V=k^{2}_{1}+\frac{1}{2}g^{2}\xi^{2}$. As
we will see, the $R$-symmetric gauge mediation with such a hidden
sector has distinctive phenomenologies compared with only $F$ term
induced SUSY-breaking \cite{Poppitz}.

For simplicity, we have assumed that the Kahler potential is
canonical, i.e,
$K=X_{n}^{\dagger}X_{n}+\Phi_{m}^{\dagger}\Phi_{m}$, In principle,
it can be modified by the radiative corrections which can in turn
induce a non-trivial metric of moduli space.

\section{A concrete example of hidden sector}

Consider a hidden sector with the gauge groups of $SU(5) \times
U(1)$, two types of chiral superfields $Q$ in the fundamental
representation, and two types of chiral superfields $\bar Q$ in
the anti-fundamental representation, as shown in the table. The
global flavor symmetry is $SU(6)$. The $U(1)_{H}$ charges of $Q$'s
and $\bar Q$'s are either $+1$ or $-1$, so the model is anomaly
free.
\begin{table}
\begin{center}
\begin{tabular}{|c|c|c|}
  \hline
   & $SU(5)$ & $U(1)_{H}$ \\
  \hline
  $Q_{i}$ & $\square$ & $+1$ \\
  $\bar{Q}_{i}$ & $\bar{\square}$ & $-1$ \\
  $Q_{6}$ & $\square$ & $0$ \\
  $\bar{Q}_{6}$ & $\bar{\square}$ & $0$ \\
  \hline
\end{tabular}
\caption{$Q_{i}$ denote the first five chiral superfields. Note
that the gauge groups in the hidden sector are the simplest
extension of that in \cite{Poppitz}, which are well motivated and
can be easily constructed in intersecting branes models.}
\end{center}
\end{table}

In the infrared, the strong coupling $SU(5)$ theory can be
described by the dual magnetic theory of ISS-type \cite{ISS}, with
a superpotneial
\begin{equation}
 W_{mag}=\lambda\bar{q}\mathcal{M}q-f^{2}Tr\mathcal{M}
\end{equation}
where $q=(\varphi,\psi)$ are the dual quarks and $\mathcal{M}$ the
mesons,
\begin{equation}
\mathcal{M}=\left(%
\begin{array}{cc}
  \omega M & kN \\
  k\bar{N} & k'Y \\
\end{array}%
\right)
\end{equation}
In detail, the superpotential of our theory can be written
as\footnote {We can also begin with a hidden sector with
superpotential $W_{mag}$ and $D$-term, and consider the
$SU(5)\times U(1)_{H}$ theory as one ultraviolet completion. For
instance, abelian SUSY theory with a set of chiral superfields
including singlets is another candidate.},
\begin{eqnarray}{\label{superpotential}}
W_{mag}&=&X^{n}f_{n}(\Phi_{m})\nonumber\\
f_{1}&=&-f^{2}\omega+\lambda\varphi\bar{\varphi},\nonumber\\
f_{2}&=&\lambda k\bar{\varphi}\psi,\nonumber\\
f_{3}&=&\lambda k\varphi\bar{\psi},\nonumber\\
f_{4}&=&-f^{2}+\lambda k'\bar{\psi}\psi,
\end{eqnarray}
where $X_{n}$ denote the mesons singlets $(M,N,\bar{N},Y)$ of
$R$-charge 2, the rest of $R$-charge 0. When the $U(1)_H$ coupling
constant $g\rightarrow 0$, the theory returns to the usual Seiberg
duality \cite{duality} and corresponding ISS model.

Here is the rational for such an assumption. In the standard model
the strong $SU(3)_{c}$ quark theory can be well described by the
dual effective theory of composite mesons and baryons at low
energy, which are not invalidated by extra electroweak
interactions. To reach the macroscopic superpotential
\eqref{superpotential}, we have assumed that the $U(1)_{H}$ gauge
theory does not spoil the validity of Seiberg duality with small
enough coupling constant $g$. There are spontaneously broken $N=2$
dual theories with FI terms \cite{IZ}, and $N=1$ dualities which
are not spoiled by deformations of IR irrelevant couplings
\cite{deformation}. There are also Seiberg dualities with
non-simple Lie groups. For example, there are $N=1$ theories with
two gauge groups \cite{GSD}. Similar assumption has been applied
to study other topics in earlier works \cite{W}.

We now read out the $U(1)_{H}$ charges of the dual mesons and
quarks. The singlet mesons $M,~ Y$ are also $U(1)_{H}$ singlets .
The $N,~ \bar{N}$ mesons are not $U(1)_{H}$ singlets, but of
$U(1)_{H}$ charges $+1,~ -1$, respectively. The dual quarks $q$
carry the same $U(1)_{H}$ charges as those of $Q$. In summary, all
dual fields are $U(1)_{H}$ singlets except
$\varphi,\bar{\varphi},N,\bar{N}$.

It can be shown that the supersymmetry is broken in our model. The
absolute minimum of $V$ is located at the origin of muduli space
with
\begin{eqnarray}{\label{values}}
F_{Tr~M}=\omega
f^{2},~~~~~\bar{\psi}\psi=\upsilon^{2}=f^{2}/(\lambda k')
\end{eqnarray}
Since $\psi,~ \bar{\psi}$ have no $U(1)_{H}$ charges, their
nonzero vacuum expectation values do not break the $U(1)_{H}$
gauge symmetry spontaneously\footnote {Recently, it is proposed
that unbroken weak coupling $U(1)_{H}$ theory can work as a model
of dark matter \cite{Weniger}.}. However, the global $SU(6)$
symmetry is spontaneously broken to $SU(5)$. The corresponding
Nambu-Goldstone (NG) bosons acquire significant masses due to
interactions with the messenger sector \cite{Poppitz}.

To mediate the SUSY-breaking to the visible sector, we gauge the
remaining global $SU(5)$ flavor symmetry. The $\phi$ and $N$
fields will serve as the messengers. The scalar mass matrix for
messengers is given by,
\begin{eqnarray}
\left(%
\begin{array}{cccc}
 \varphi^{*} & \bar{\varphi} & N^{*} & \bar{N} \\
\end{array}%
\right)
\left(%
\begin{array}{cccc}
  M^{2}+D & -zM^{2} & 0 & 0 \\
  -zM^{2} & M^{2}-D & 0 & 0 \\
  0 & 0 & M^{2}+D& 0 \\
  0 & 0 & 0 & M^{2}-D \\
\end{array}%
\right)
\left(%
\begin{array}{c}
  \varphi \\
  \bar{\varphi}^{*} \\
  N \\
  \bar{N}^{*} \\
\end{array}%
\right) \label{massmatrix}
\end{eqnarray}
where $M=\sqrt{\lambda\omega/z}f$ and $z=\omega k'/k^{2}$. From
Eq. (\ref{massmatrix}) we see that the eigenstates of the upper
$2\times2$ block of scalar mass matrix are
$\phi_{\pm}=(\bar{\varphi}^{*}\pm \varphi)/\sqrt{2}$ of
eigenvalues,
\begin{eqnarray}{\label{M}}
\phi:~~ m_{\pm}^{2}=M^{2}(1\pm \tilde{z}),
~~~~~~\tilde{z}=\sqrt{z^{2}+x^{2}}
\end{eqnarray}
where $x=D/ M^{2}$. In order to avoid tachyons in the spectrum, we
need to impose $x^{2}<(1-z^{2})$. The eigenvalues of $N,\bar{N}$
are given by,
\begin{eqnarray}{\label{N}}
N:~~ m_{\pm}^{2}=M^{2}(1\pm x)
\end{eqnarray}
The fermion mass matrix for messengers $\varphi, N$ is
off-diagonal,
\begin{eqnarray}
\left(%
\begin{array}{cc}
  \varphi & N \\
\end{array}%
\right)
\left(%
\begin{array}{cc}
  0 & Me^{i\theta} \\
  Me^{-i\theta} & 0 \\
\end{array}%
\right)
\left(%
\begin{array}{c}
  \bar{\varphi} \\
  \bar{N} \\
\end{array}%
\right)
\end{eqnarray}
They are all degenerate at $M$. The spectra for NG particles and
the remaining messengers $X,\psi,\bar{\psi}$ are the same as those
given in \cite{Poppitz}. Compared with the messenger spectra in
\cite{Poppitz}, the masses of scalar messengers $\phi$ and $N$ are
modified by the $D$-term. The sfermions masses squared and Dirac
gaugino masses in the visible sector will be induced by the mass
splitting of $\phi$ and $N$ in the loop(s).

\section{Soft terms and fine tunings}
We now analyze the soft terms in the visible sector. Before going
to the details, we outline the main characteristics of
$R$-symmetric gauge mediation with $D$-term,
\begin{itemize}
    \item The sfermion masses are decreased by negative $D$-term induced contribution.
When $\sqrt{D}\sim M\sim 10^{3}$TeV, the sfemions are still of
order $\mathcal{O}(1)$TeV. On the other hand, the Dirac gaugino
masses increase by positive $D$-term induced contribution. Without
$D$-term effects the sfermions masses are usually heavier than
gauginos masses. However, this can easily be reversed with the
$D$-term. This reversion of the gaugino and scalar mass ratio
helps to evade constraints on flavor structures.
    \item As shown in \cite{Poppitz}, fine tunings are needed to have viable diagonal and off-diagonal
coefficients $c_{D}$ and $c_{OD}$ for the sfermion mass matrix. An
adjustable $D$-term helps to obtain reasonable relations
$c_{D}\sim c_{OD}\sim 1$.
    \item $B\mu$ and $\mu$ terms receive the nonzero and zero contributions from $D$-term respectively,
which can be used to adjust the small hierarchy between $\mu$ and
$B\mu$.
    \item Gravitino is the lightest superparticle (LSP) of a mass in the order of $eV$,
which is determined by $m_{3/2}\sim (D+f^{2})/M_{Pl}$ in
supergravity.
\end{itemize}
The sfermion masses squared receive contributions from the
ultraviolet (UV) and infrared (IR) physics. The IR contribution is
due to the mass splitting of the messengers induced by the SUSY
breaking in the hidden sector, starting from two-loops
\cite{Martin}. The UV contribution can be written generically,
\begin{eqnarray}
\int d^{4}\theta
\frac{c_{ij}}{\Lambda^{2}}(\Xi\Xi^{\dagger})Q^{\dagger}_{i}Q_{j}
\end{eqnarray}
where $\Xi=<TrM>=\theta^{2}\omega f^{2}$. In total, we have
\begin{eqnarray}{\label{sfermion mass}}
(\tilde{m}^{2})_{ij}&=&c_{ij}(\tilde{m}_{UV}^{2})_{ij}-(\tilde{m}_{IR}^{2})_{ij},\nonumber\\
(\tilde{m}_{IR}^{2})_{ij}&=& \delta_{ij}\frac{g_{s}^{4}M^{2}}{(16\pi^{2})^{2}}J(x,\tilde{z}),\nonumber\\
\tilde{m}_{UV}&=&\left(\frac{z}{\lambda}\right)\left(\frac{M}{\Lambda}\right)M
\end{eqnarray}
where
\begin{eqnarray}{\label{J}}
J(x,\tilde{z})=\frac{7}{9}(x^{4}+\tilde{z}^{4})+\frac{38}{75}(x^{5}+\tilde{z}^{5})
+\mathcal{O}\left(x^{6},\tilde{z}^{6}\right)
\end{eqnarray}
$c_{ij}$ are the coefficients appearing in the UV operator. The UV
operators responsible for Dirac gaugino masses are,
\begin{eqnarray}
\int d^{2}\theta
\frac{\bar{D}^{2}D_{\alpha}(\Xi\Xi^{\dagger})}{\Lambda^{3}}Tr\left(W^{\alpha}\Phi\right)
\end{eqnarray}
where $W',W$ refer to the $U(1)_{H}$ and SSM spinor superfields
respectively. The IR contribution to gaugino masses is again due
to the mass splitting of the messengers, starting from one-loop.
Since the masses of the scalars $N,\bar{N}$ in the loop are
shifted, gaugino masses are changed, in comparison with those in
\cite{Poppitz}. Explicitly, we have
\begin{eqnarray}{\label{gaugino mass}}
m_{1/2}&=&m_{IR}+m_{UV},\nonumber\\
m_{IR}&=&\frac{g_{s}y}{16\pi^{2}}Mcos\left(\frac{\theta}{\upsilon}\right)Q(x,\tilde{z})
\end{eqnarray}
where
\begin{eqnarray}{\label{gaugino UV}}
m_{UV}\simeq
\left(\frac{\tilde{m}_{UV}}{\Lambda}\right)\tilde{m}_{UV}
\end{eqnarray}
and $\theta$ is the NG bosons, The function $Q(x,\tilde{z})$ is
defined by ,
\begin{eqnarray}
Q(x,\tilde{z})&=&\frac{1}{\tilde{z}}\left((1+\tilde{z})\log(1+\tilde{z})-(1-\tilde{z})\log(1-\tilde{z})-2\tilde{z}\right)
+(\tilde{z}\rightarrow x)
\end{eqnarray}
The coefficients $c_{ij}$ can also be written as \cite{Poppitz},
\begin{eqnarray}
c_{D}=\frac{\tilde{m}^{2}+\tilde{m}^{2}_{IR}}{\tilde{m}^{2}_{UV}},~~~~~~~
c_{OD}=\delta\left(\frac{\tilde{m}^{2}}{\tilde{m}^{2}_{UV}}\right)
\end{eqnarray}
which refer to the non-perturbative behavior of the hidden sector,
which arises when the RG scale near the Landau pole $\Lambda$ in
the direct gauge mediation\footnote{ The messengers introduced to
mediate the SUSY-breaking effects substantially modify the slopes
of gauge coupling $\alpha_{s}$ running, lead to the divergence of
$\alpha_{s}$ at $\Lambda$.}. In principle, $c_{ij}$ are of the
order $\mathcal{O}(1)$ and cannot be calculated in perturbative
method. However, they are constrained by ratio of gaugino mass
over sfermion mass to avoid flavor problems. In the ISS model,
$c_{D}$ should be smaller than $10^{-2}$ to be phenomenlogically
viable. This implies that the non-perturbative physics of the
hidden sector is seriously constrained and this is the origin of
fine tuning.

In models with $D$-term breaking, $c_{D}$ can be in the
neighborhood of unity. Take $M\sim 10^{3}$TeV and $\Lambda\sim
10^{4}$TeV for illustrations, in which $\alpha_{3}(M)\sim 0.15$.
We start with the following parameter space,
\begin{eqnarray}{\label{parameter}}
z=0.1~~~ and~~~~\lambda=1,
\end{eqnarray}
which can yield the typical mass relations,
\begin{eqnarray}{\label{mass relation}}
\tilde{m}\sim \tilde{m}_{UV}\sim \tilde{m}_{IR}\sim 10^{-3}M,~~~~
m_{1/2}\sim m_{IR}\sim 10^{-2}M
\end{eqnarray}
Explicit calculations show that $c_{D}\leq 10$ and $c_{OD}\leq 1$
are phenomenlogically viable. Shown in Figure. 1 are the masses of
sfermions and gauginos for two typical regions of $c_{ij}$.
Clearly, for the given parameter space \eqref{parameter}, there
are no fine tunings and no extra Landau pole coming from large
Yukawa coupling $\lambda$.

When going up along the positive direction of $z$, one has to
increase $x\sim 1$ and (or) $\lambda$ in order to restore the
typical mass relations \eqref{mass relation}. However, there is
the upper bound on $x\sim (1-z)$, as one wants to avoid tachyons.
Roughly $z$ cannot be greater than $0.5$.

\begin{figure}
\centering
\begin{minipage}[b]{0.5\textwidth}
\centering
\includegraphics[width=3in]{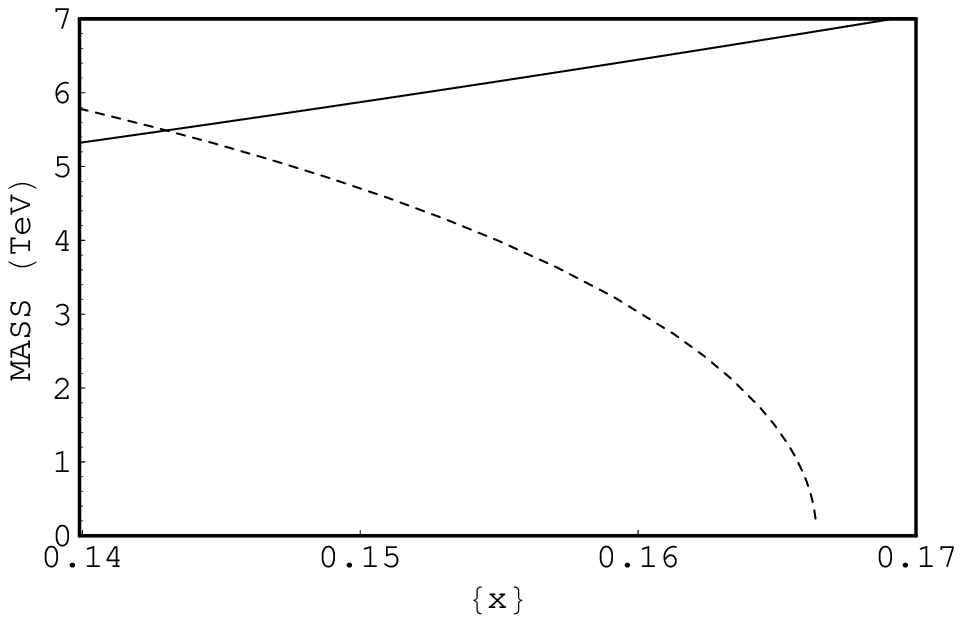}
\end{minipage}%
\begin{minipage}[b]{0.5\textwidth}
\centering
\includegraphics[width=3in]{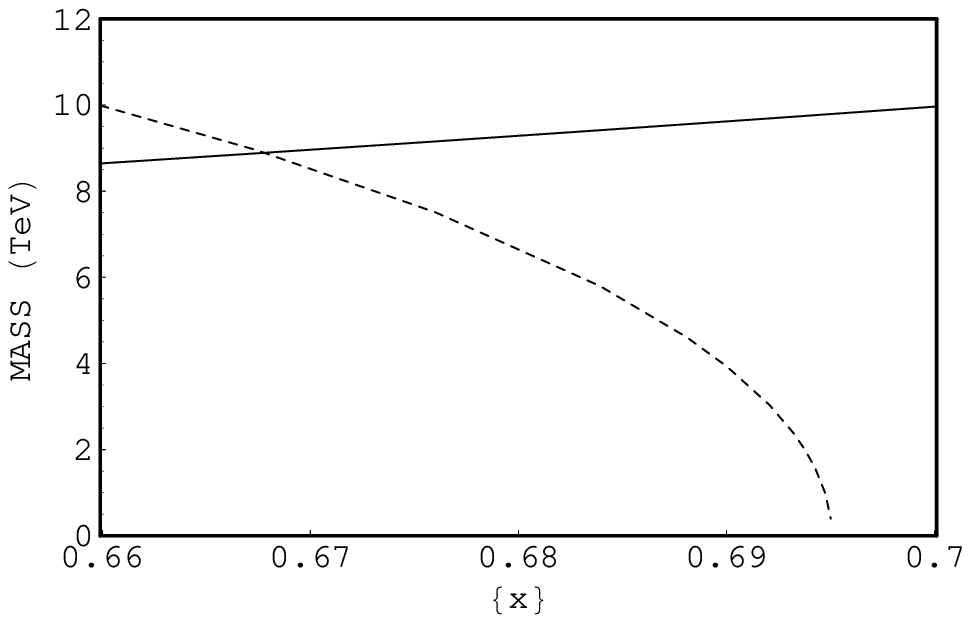}
\end{minipage}
\caption{ The massses of sfermion (dashed line) and gaugino (solid
line) (TeV scale) as functions of $x$ parameter. Left figure is
for the region of small $c_D=0.1$, where a large mass ratio can be
obtained near $x\sim 0.16$ ($y=4$ and $M=3\times10^{3}$ TeV).
Right figure is for the region of large $c_D=8$, where a large
mass ratio can be obtained near $x\sim 0.69$ ($y=3$ and $M=10^{3}$
TeV).}
\end{figure}

Finally, we address the $\mu/B\mu$ term in $R$-symmetric gauge
mediation. Generally, there are three mechanisms to generate
$\mu/B\mu$ term of weak scale in gauge mediation without
$R$-symmetry. One is by introducing a gauge singlet $X$, usually
dubbed as NMSSM \cite{Giudice97}. Another is by introducing
massive vector-like pairs. Thirdly, one takes the conformal
sequestering into account \cite{Murayama}. In $R$-symmetric gauge
mediated theories, the unbroken $R$-symmetry severely restricts
the $\mu$ term, while $B\mu$ of weak scale can be generated from
either the $F$- or $D$-terms. An economical scheme to the $\mu$
term is to introduce two extra $SU(2)$ doublets $R_{u,d}$ of
$R$-charge 2 \cite{Poppitz}. The UV operators corresponding to
$\mu$  and $B\mu$ are
\begin{eqnarray}{\label{u}}
\int d^{4}\theta
c_{F}\frac{\Xi^{\dagger}}{\Lambda}\left(H_{u}R_{d}+
H_{d}R_{u}\right)
\end{eqnarray}
and,
\begin{eqnarray}{\label{Bu}}
\int d^{4}\theta
c'_{F}\frac{\Xi\Xi^{\dagger}}{\Lambda^{2}}H_{u}H_{d} +\int
d^{2}\theta c'_{D}\frac{W'_{\alpha}W'^{\alpha}}{M^{2}}H_{u}H_{d}
\end{eqnarray}
respectively. Here the second term comes from the unbroken
$U(1)_{H}$ gauge theory. They are both at the weak scale if
$\sqrt{D}\sim M\sim 10^{3}$TeV
\begin{eqnarray}
\mu\sim c_{F}m_{UV},~~~~ B\mu\sim
c'_{F}m_{UV}^{2}+c'_{D}\frac{D^{2}}{M^{2}}
\end{eqnarray}

Note that the simplest way to generate $\mu$ term is to include a
$R$-charge zero spurion field of nonzero $F$-term in the messenger
sector. According to discussions in section 2, this can be
realized. For instance, the Wess-Zumino model with only chiral
superfields is a feasible candidate. It would be interesting to
develop a scheme to dicuss the general $R$-symmetric gauge
mediation as done in ordinary gauge mediation \cite{MSS}, to see
which of the features outlined above are general and which are
model dependent.

\section{Conclusions}
In this paper, we have discussed the possibilities to obtain
SUSY-breaking hidden sectors with $R$-symmetry, with an extra
$U(1)_H$ sector and a FI term. A concrete example of hidden sector
is constructed. In this particular model, we find that in the
visible sector, the ratio of Dirac gaugino mass over sfermion mass
substantially increases compared with those with only $F$-term
\cite{Poppitz}. The $\mu$ and $B\mu$ terms receive zero and
nonzero contribution, respectively. These help evading the fine
tunings in $R$-symmetric gauge mediation with interesting flavor
phenomenologies.

As we point out in section 2, there are other candidates as hidden
sectors in $R$-symmetric gauge mediation. It is worth developing a
general framework to see which characters are model independent,
especially the generations of $\mu/B\mu$ terms, which closely
connect with some important topics such as electro-weak symmetry
breaking and the dark matter model of supersymmetric neutralino.

\section*{Acknowledgement}
This work is supported in part by the National Science Foundation
of China (10425525).

\end{document}